# Investigation on non-ergodicity of protein dynamics

*Luca Maggi*

*NBD | Nostrum Biodiscovery, Josep Tarradellas 8-10 Barcelona 08019, Spain.*

luca.maggi@nostrumbiodiscovery.com

## Abstract:

The study of microscopic protein dynamics has historically presented significant challenges to researchers seeking to develop a comprehensive and detailed description of its diverse and intriguing features. Recent experimental and theoretical studies have proposed the hypothesis that protein dynamics may be non-ergodic. The implications of this finding are of paramount importance from both a practical and theoretical standpoint. In this study, we employ all-atom molecular dynamics simulations to examine these results over a time window spanning from picoseconds to nanoseconds. To this end, we utilize widely used statistical tools. Our findings challenge the conclusions of previous studies, which suggested that proteins exhibit non-ergodic dynamics. Instead, we demonstrate that deviations from ergodic behavior are due to incomplete convergence of the investigated quantities. Additionally, we discuss the implications of findings that suggest a potential breaking of the ergodic hypothesis over larger time windows, which were not directly investigated in this study.

## Introduction:

Proteins are highly dynamic entities. In order to perform their biological function, they must be able to adopt a number of different conformations[1]. These involve spatial rearrangements of their constitutive elements at varying length and time scales including short and rapid residue rotations as well as large and slow displacement of an entire domain. These events are closely intertwined in an extremely convoluted and hierarchical order[2,3] which gives rise to a highly complex dynamic. This presents strikingly interesting physical properties including the excess of low frequency modes [4,5](i.e. the "boson peak"), non-Arrhenius[6–8] laws governing phenomena and the reduced diffusion in the conformational space[9,10]. All of them directly impact on biologically protein functions as, for instance, the enzyme activity[6,7]. Significant resources have been dedicated to developing theoretical models to elucidate the diverse characteristics of protein dynamics[11–16]. These models exhibit remarkable differences due to their disparate underlying assumptions and the properties they highlight. However, most of them assume that protein dynamics is ergodic. In general, ergodicity refers to the equality between the ensemble and the time average of an infinite-long trajectory[17]. This equality is contingent upon the system's capacity to explore the entirety of the available phase space. Nevertheless, there are instances in which the comprehensive sampling of the phase space cannot be accomplished within the pertinent time scale of the system, thereby violating the principle of ergodicity. This phenomenon can be observed, for instance, in glassy systems[18]. Recent studies, inspired by the parallels between protein and glassy dynamics[3,8,19–21], have presented novel experimental and theoretical evidence suggesting that protein dynamics may be non-ergodic[13,22–24]. The multitude of metastable states with disparate exit times renders the exploration of the conformational state particularly sluggish, akin to a viscous flow. Protein dynamics is, thus, perpetually aging and non-equilibrium, with an incomplete sampling that ultimately results in an ergodicity breaking[13,24]. These studies assume a crucial importance in the context of all-atom molecular dynamics (MD) simulations. The results of MD simulations indeed are typically derived from relatively few long trajectories, which are then compared with experimental outcomes represented by ensemble averages[25,26]. This comparison is justified from a formal standpoint by the assumption of system ergodicity. If this principle is no longer applicable, the theoretical foundation for the comparison will also be rendered invalid. However, these recent works are rather heterogeneous focusing on disparate time windows, spanning from picoseconds[24] to hours[23] and showing puzzling effects including the possibility that the ergodicity breaking may be contingent upon the number of degrees of freedom inherent to the protein under examination. It appears that smaller systems continue to adhere to the ergodic principle, whereas larger proteins do not[24]. In light of these considerations, the broad relevance and the fundamental question posed by these studies, as well as the necessity to depict a clear picture of this phenomenon, demand a thorough scrutiny to confirm and clarify these remarkable findings. In this work, we will employ MD simulations to verify the ergodicity-breaking hypothesis in the pico-second to nano-second range. To achieve this objective, we examine the dynamics of two distinct proteins and calculate relevant quantities commonly used to assess ergodicity.

## Methods and Results:

The proteins selected for this work as case studies are as follows: A 35-residue sequence of the Villin headpiece, i.e. the binding domain located at the C-terminus of the F-actin protein, whose activity is crucial for microvilli formation and which will be referred to as Villin in the following text (PDB ID: 1D5G[27]), and the catabolite gene activator protein (CAP) in complex with cyclic AMP (PDB ID: 1G6N[28]), which is responsible for the activation of a large number of genes. Binding of cyclic AMP induces an allosteric conformational change that allows CAP to bind a specific DNA sequence. The two proteins exhibit notable differences in their respective sizes. The former is slightly longer than a polypeptide, comprising 35 residues, while the latter is a homodimer comprising approximately 400 residues. This discrepancy will assist us in elucidating the enigmatic phenomenon observed by Li et al., which suggests that the ergodicity of biomolecules may be contingent upon their size[24]. Consequently, 100 independent 100 $ns$-long molecular dynamics (MD) simulations were conducted on these systems from which a range of physical quantities were extracted to describe their dynamics. The starting configuration of each MD replica is extracted from a 1 $\mu s$- long simulation. Details on how MD simulations were conducted are described elsewhere[12]. In the case of Villin, two descriptors will be employed: the root mean square difference (RMSD) of the $\alpha$-carbons from a reference conformation of two specific segments of the protein, as illustrated in Fig. 1. These have previously been employed to effectively describe the dynamics of this small protein[29]. In contrast, the CAP dynamics is defined by a single quantity: the distance between the two DNA-binding domains. Specifically, the distance between the center of mass of two arginine residues located on the surface of these domains was measured (see Fig. 1). The diversity of the physical descriptors for the dynamics will assist us in determining whether the non-ergodicity may arise exclusively when associated with a specific set or number of variables. By employing these descriptors, we can assess the ensemble averaged mean square displacement (MSD), defined as:

$$\langle \Delta X^2(t) \rangle = \int dX\, dX(0)\, \rho(X,\, t;\, X(0), t=0)\, |X(t) - X(0)|^2 \qquad \text{Eq. 1}$$

Where $X(t)$ is the values of the RMSD of the two segments (i.e. a vector) in case of Villin or the distance between the two domains for CAP at time $t$. $\langle ... \rangle$ denotes the ensemble average and $\rho(X, t; X(0), t=0)$ is the joint probability distribution function of the variable $X$ at time $t$ and $X(0)$ at $t=0$. Since $\rho$ is not directly accessible, the evaluation of MSD is achieved here by averaging $|X(t) - X(0)|^2$ over all the replicas:

$$\langle \Delta X^2(t) \rangle = \frac{1}{N} \sum_k^N |X_k(t) - X_k(0)|^2 \qquad \text{Eq. 2}$$

Where $N$ is the total number of replicas and $X_k$ is the chosen physical descriptor of the $k$-th replica. The time-dependent mean square displacement (TA-MSD) of single simulation $k$ is evaluated using a moving average as follow:

$$\overline{\delta_k^2(T,t)} = \frac{1}{T-t} \int_0^{T-t} d\tau \, |X_k(t+\tau) - X_k(\tau)|^2 \qquad \text{Eq. 3}$$

Where $T$ is the length of each replica and $t$ is the so-called lag-time. Eq.2 and 3 define two central quantities of this study since the ergodic hypothesis can be mathematically stated as the equality between them[17]:

$$\langle \Delta X^2(t) \rangle = \lim_{T/t \to \infty} \overline{\delta_k^2(T,t)} \qquad \text{Eq. 4}$$

This should be true for all $k$. In case the Eq.4 doesn't hold the system is considered non-ergodic. Usually, the ergodic hypothesis is simply formulated as $T \to \infty$, but since we are dealing with finite time windows, it is necessary to work with relative nondimensional quantities, where it is easier to identify a threshold defining the "high values" regime[30]. Based on previous work on sub-diffusion in proteins[12,15], this is identified when $T$ is at least two orders of magnitude larger than the lag time. Therefore, since $T$ is $100 \, ns$, $t$ never exceeds $1 \, ns$. Despite the direct comparison between the ensemble and time-averaged MSD as in Eq. 4 appearing to be the most straightforward method for evaluating the ergodicity of the systems, this procedure does not yield reliable results. Indeed, $\overline{\delta_k^2(T,t)}$ exhibits considerable variability among different replicas, as illustrated in Fig. 2. This renders the estimation of equality in Eq. 4 highly inaccurate. This variability should not be interpreted as a sign of non-ergodicity, as it may result from an incomplete convergence of the calculated quantities due to the short length of measurements. If this were the case, the variability would be expected to disappear upon increasing the length of the simulation. Conversely, it should persist, at least within the time window pertinent with protein function, in the event of a non-ergodic system[30]. Therefore, analyzing how this variability behaves as $T/t \to \infty$ can provide us crucial insight on the system ergodicity. To this aim we first define the ensemble average of $\overline{\delta_k^2(T,t)}$ as:

$$\langle \overline{\delta^2(T,t)} \rangle = \frac{1}{N} \sum_k^N \overline{\delta_k^2(T,t)} \qquad \text{Eq. 5}$$

Which is instrumental to introduce a widely used quantity to study the trajectory-to-trajectory variability; the ergodicity breaking ($EB$) parameter defined as follow[31,32]:

$$EB(T,t) = \frac{Var\ \overline{\delta_k^2(T,t)}}{\langle\overline{\delta^2(T,t)}\rangle^2} \qquad \text{Eq.6}$$

Where $Var\ \overline{\delta_k^2(T,t)}$ is the variance of $\overline{\delta_k^2(T,t)}$ which is, as usual, defined as $\langle\left(\overline{\delta_k^2(T,t)} - \langle\overline{\delta^2(T,t)}\rangle\right)^2\rangle$. Moreover, after Introducing the parameter $\xi_k(T,t) = \frac{\overline{\delta_k^2(T,t)}}{\langle\overline{\delta^2(T,t)}\rangle}$ and simple mathematical manipulations Eq.6 can be recast as[30]:

$$EB(T,t) = \langle(\xi_k(T,t) - 1)^2\rangle \qquad \text{Eq.7}$$

For ergodic system which exhibit very low trajectory-to-trajectory variability, the variance of $\overline{\delta_k^2(T,t)}$ becomes negligible with respect its squared average as $T/t \to \infty$, thus $EB$ converges to zero. A second relevant quantity, related to $EB$, is the distribution $\phi$ of $\xi_k(T,t)$. In case it converges to a delta distribution centered on 1, the TA-MSD of each replica coincides with its ensemble average $\langle\overline{\delta^2(T,t)}\rangle$ indicating the ergodicity of the system[30]. Therefore, whether those presented quantities indicate a reduced variability between $\overline{\delta_k^2(T,t)}$ and $\langle\overline{\delta^2(T,t)}\rangle$ we obtain a strong suggestion of the ergodicity and moreover we are allowed from a theoretical standpoint to compare the $\langle\overline{\delta^2(T,t)}\rangle$, instead of $\overline{\delta_k^2(T,t)}$, to the ensemble average $\langle\Delta\mathbf{X}^2(t)\rangle$ to evaluate directly the ergodicity statement.

Fig. 3 illustrates the monotonic decrease of the $EB$ parameter in conjunction with $T$ and varying lag times $t$ for both the investigated proteins. The $EB$ is not evaluated for the entire range of $T$ values; rather, its lower limit is set by the selected $t$ value to maintain its ratio larger than $10^2$ as noted above. The $EB$ parameter consistently exhibits minimal values, remaining below $10^{-1}$. According to Eq. 6, thus, the variance of TA-MSD is at least one tenth of $\langle\overline{\delta^2(T,t)}\rangle^2$. Hence, the trajectory-to-trajectory variability can be considered significantly reduced in each protein studied, and it decreases as $T/t \to \infty$. This is a typical feature of an ergodic system. This finding is wholly corroborated by an examination of the $\phi$ distribution profiles, as illustrated in Fig. 4. It is evident that the distribution is centered at 1 and exhibits a narrow Gaussian-shaped profile akin to a delta distribution. The calculation was performed with $T = 100$ ns and varying lag times. As with the $EB$ parameter, the ratio between these two values is always larger than $10^2$. As shown in Fig. 4, the variance of the distribution, which corresponds exactly with the EB parameter since the average value is 1, increases as the lag time increases. Nevertheless, the shape is maintained as well as the mean value. Additionally, the distribution presents a null value at zero, whereas typical non-ergodic models exhibit values greater than zero at the origin[33,34]. This finding provides further evidence of the reduced variability and supporting the idea that ergodicity is not violated within the investigated time range. As we have demonstrated the reduced statistical variability of TA-MSDs as definitive evidence of system ergodicity, we directly compare $\langle\Delta\mathbf{X}^2(t)\rangle$ with $\langle\overline{\delta^2(T,t)}\rangle$. Fig. 5 illustrates that the profiles of these two quantities are quite similar. To quantify this similarity, we calculated the root-square mean square error, normalized by the difference between the maximum and minimum values. This yielded small values for both protein that are 15% for the Villin profile and 11% for the CAP profile. The noise in the $\langle\Delta\mathbf{X}^2(t)\rangle$ profile is primarily attributed to the number of simulated replicas which will disappear increasing $N$.

## Discussion:

The results of the MD simulations clearly demonstrate that protein dynamics is ergodic in the pico-to-nanosecond range, regardless of the size of the molecule under study. However, on larger time scales, experimental studies indicate that ergodicity may be broken due to the lack of convergence of the investigated dynamical descriptors after prolonged and repeated measurements. By employing a plasmon resonance-based technique, Ye et al. have succeeded in monitoring a single protein conformational change for an unprecedented duration[23]. It is noteworthy that the authors reported an exceptionally prolonged conformational transition time, estimated to be in the range of minutes. While this observation offers intriguing insights into protein dynamics, it does not constitute definitive evidence of non-ergodicity. Conversely, the observed heterogeneity in the percentage of residence time in each conformation, derived from multiple independent long realizations, could be indicative of a lack of ergodicity.Similarly, Li et al. in documented remarkable heterogeneity in multiple independent smFRET experiments[24,35]. The assessment of apparent FRET efficiency reveals the existence of multiple populations, whose transitions are not observed in individual measurements. Consequently, it has been proposed that the protein is unable to fully explore the conformational space accessible within the time frame associated with its function. This kind of heterogeneity is already known as "static heterogeneity"[35,36]. It refers to samples that include species that cannot dynamically interchange with each other, at least within the measurement time window. If we consider this heterogeneity to be a specific feature of protein dynamics, then each conformational state observed in the experiments should be regarded as an independent protein *native state*, associated with different dynamical descriptors. The inability to observe the dynamical transition implies that the protein must assume each different conformational state during the folding process which, instead of being commonly thought as a "funnel" process leading to a single native state, should include multiple funnels associated to different folded conformation. The high energy barriers separating these conformations define, thus, disjoint valleys in the conformational space, each valley includes several metastable states among which the dynamics of a folded protein occurs. Our results clearly show that this dynamic is "locally ergodic" within a valley, however, it is *by definition* non-ergodic as a single protein molecule cannot change its native state by visiting another valley. Therefore, there should be a clear separation between the exiting times associated to each metastable within each valley and those time related to escaping from a valley, which should be beyond the functional lifetime of a folded protein. This separation produces a discontinuous distribution of the exiting-times in partial disagreement with the idea put forward by Xe et al.[13], according to which the microscopic model gives rise to non-ergodicity may be a continuous time random walk with power-law distributed exiting times that produces an average exiting time much larger than the time pertaining to protein function. More importantly, the multi-funnel scenario is at odds with the structural biology paradigm, which associates to a specific sequence a particular folded structure[37]. Although this structure is highly dynamic and adopts multiple conformations the native state is always found to be unique[1]. This raises questions about the origin of this observed heterogeneity, which may alternatively originate from the experimental conditions and setup.In conclusion, our findings demonstrate that protein exhibits a completely ergodic dynamic within a time window ranging from pico-to-nanoseconds, irrespective of their size. This result has been achieved by employing statistical quantities that are commonly utilized to assess ergodicity, calculated from the outcomes of molecular dynamics simulations. We do believe that further investigation is required to elucidate and understand this phenomenon on larger time scales.

## Reference:

1.  Henzler-Wildman, K. & Kern, D. Dynamic personalities of proteins. *Nature* **450**, 964–972 (2007).

2. Frauenfelder, H., Sligar, S. G. & Wolynes, P. G. The Energy Landscapes and Motions of Proteins. *Science (1979)* **254**, 1598–1603 (1991).

3. Iben, I. E. T. *et al.* Glassy behavior of a protein. *Phys Rev Lett* **62**, 1916–1919 (1989).

4. Mori, T. *et al.* Detection of boson peak and fractal dynamics of disordered systems using terahertz spectroscopy. *Phys Rev E* **102**, 022502 (2020).

5. Nakagawa, H., Joti, Y., Kitao, A., Yamamuro, O. & Kataoka, M. Universality and Structural Implications of the Boson Peak in Proteins. *Biophys J* **117**, 229–238 (2019).

6. Flomenbom, O. *et al.* Stretched exponential decay and correlations in the catalytic activity of fluctuating single lipase molecules. *Proceedings of the National Academy of Sciences* **102**, 2368–2372 (2005).

7. English, B. P. *et al.* Ever-fluctuating single enzyme molecules: Michaelis-Menten equation revisited. *Nat Chem Biol* **2**, 87–94 (2006).

8. Morgan, I. L., Avinery, R., Rahamim, G., Beck, R. & Saleh, O. A. Glassy Dynamics and Memory Effects in an Intrinsically Disordered Protein Construct. *Phys Rev Lett* **125**, 058001 (2020).

9. Yang, H. *et al.* Protein Conformational Dynamics Probed by Single-Molecule Electron Transfer. *Science (1979)* **302**, 262–266 (2003).

10. Grossman-Haham, I., Rosenblum, G., Namani, T. & Hofmann, H. Slow domain reconfiguration causes power-law kinetics in a two-state enzyme. *Proceedings of the National Academy of Sciences* **115**, 513–518 (2018).

11. Calandrini, V., Abergel, D. & Kneller, G. R. Fractional protein dynamics seen by nuclear magnetic resonance spectroscopy: Relating molecular dynamics simulation and experiment. *J Chem Phys* **133**, (2010).

12. Maggi, L. & Orozco, M. Main role of fractal-like nature of conformational space in subdiffusion in proteins. *Phys Rev E* **109**, 034402 (2024).

13. Hu, X. *et al.* The dynamics of single protein molecules is non-equilibrium and self-similar over thirteen decades in time. *Nat Phys* **12**, 171–174 (2016).

14. Ciliberti, S., Rios, P. D. L. & Piazza, F. On the origin of the boson peak in globular proteins. *Philosophical Magazine* **87**, 631–641 (2007).

15. Meroz, Y., Ovchinnikov, V. & Karplus, M. Coexisting origins of subdiffusion in internal dynamics of proteins. *Phys Rev E* **95**, 062403 (2017).

16. Kou, S. C. & Xie, X. S. Generalized Langevin Equation with Fractional Gaussian Noise: Subdiffusion within a Single Protein Molecule. *Phys Rev Lett* **93**, 180603 (2004).

17. Metzler, R. Weak ergodicity breaking and ageing in anomalous diffusion. *Int J Mod Phys Conf Ser* **36**, 1560007 (2015).

18. Thirumalai, D., Mountain, R. D. & Kirkpatrick, T. R. Ergodic behavior in supercooled liquids and in glasses. *Phys Rev A (Coll Park)* **39**, 3563–3574 (1989).

19. Chen, G., Fenimore, P. W. & Frauenfelder, H. Glassy Dynamics of Proteins. in *Structural Glasses and Supercooled Liquids* 319–339 (Wiley, 2012). doi:10.1002/9781118202470.ch9.

20. Tüzel, E. & Erzan, A. Glassy dynamics of protein folding. *Phys Rev E* **61**, R1040–R1043 (2000).

21. Teeter, M. M., Yamano, A., Stec, B. & Mohanty, U. On the nature of a glassy state of matter in a hydrated protein: Relation to protein function. *Proceedings of the National Academy of Sciences* **98**, 11242–11247 (2001).

22. Li, J. *et al.* Reply to: Insufficient evidence for ageing in protein dynamics. *Nat Phys* **17**, 775–776 (2021).

23. Ye, W. *et al.* Conformational Dynamics of a Single Protein Monitored for 24 h at Video Rate. *Nano Lett* **18**, 6633–6637 (2018).

24. Li, J. *et al.* Non-ergodicity of a globular protein extending beyond its functional timescale. *Chem Sci* **13**, 9668–9677 (2022).

25. Smith, L. J., Athill, R., van Gunsteren, W. F. & Hansen, N. Interpretation of Seemingly Contradictory Data: Low NMR $S^2$ Order Parameters Observed in Helices and High NMR $S^2$ Order Parameters in Disordered Loops of the Protein hGH at Low pH. *Chemistry – A European Journal* **23**, 9585–9591 (2017).

26. Capelli, R., Carloni, P. & Parrinello, M. Exhaustive Search of Ligand Binding Pathways via Volume-Based Metadynamics. *J Phys Chem Lett* **10**, 3495–3499 (2019).

27. McKnight, C. J., Matsudaira, P. T. & Kim, P. S. NMR structure of the 35-residue villin headpiece subdomain. *Nat Struct Biol* **4**, 180–184 (1997).

28. Passner, J. M., Schultz, S. C. & Steitz, T. A. Modeling the cAMP-induced Allosteric Transition Using the Crystal Structure of CAP-cAMP at 2.1 Å Resolution. *J Mol Biol* **304**, 847–859 (2000).

29. Lei, H., Wu, C., Liu, H. & Duan, Y. Folding free-energy landscape of villin headpiece subdomain from molecular dynamics simulations. *Proceedings of the National Academy of Sciences* **104**, 4925–4930 (2007).

30. Cherstvy, A. G. & Metzler, R. Ergodicity breaking, ageing, and confinement in generalized diffusion processes with position and time dependent diffusivity. *Journal of Statistical Mechanics: Theory and Experiment* **2015**, P05010 (2015).

31. Godec, A. & Metzler, R. Finite-Time Effects and Ultraweak Ergodicity Breaking in Superdiffusive Dynamics. *Phys Rev Lett* **110**, 020603 (2013).

32. Thiel, F. & Sokolov, I. M. Weak ergodicity breaking in an anomalous diffusion process of mixed origins. *Phys Rev E* **89**, 012136 (2014).

33. Schwarzl, M., Godec, A. & Metzler, R. Quantifying non-ergodicity of anomalous diffusion with higher order moments. *Sci Rep* **7**, 3878 (2017).

34. Mardoukhi, Y., Jeon, J.-H. & Metzler, R. Geometry controlled anomalous diffusion in random fractal geometries: looking beyond the infinite cluster. *Physical Chemistry Chemical Physics* **17**, 30134–30147 (2015).

35. Torella, J. P., Holden, S. J., Santoso, Y., Hohlbein, J. & Kapanidis, A. N. Identifying Molecular Dynamics in Single-Molecule FRET Experiments with Burst Variance Analysis. *Biophys J* **100**, 1568–1577 (2011).

36. Holden, S. J. *et al.* Defining the Limits of Single-Molecule FRET Resolution in TIRF Microscopy. *Biophys J* **99**, 3102–3111 (2010).

37. Subramaniam, S. & Kleywegt, G. J. A paradigm shift in structural biology. *Nat Methods* **19**, 20–23 (2022).

# Figures:

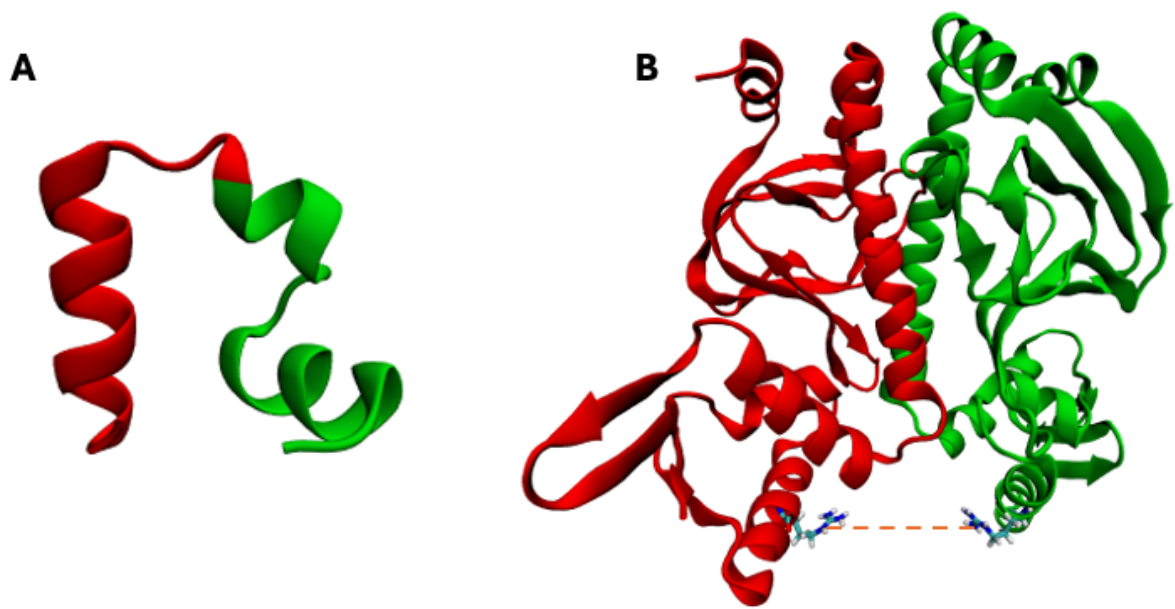

**Figure 1:** The proteins analyzed in this study. (**A**) Villin, the two segments for which the $\alpha$-carbon RMSDs have been calculated are coloured differently. The first segment comprising the first 17 residue is green while the other is red. (**B**) CAP, the two homodimers are identified with the a green and red color. The dashed orange line indicates the distance beetween the arginines used as physical descriptor of this protein dynamics.

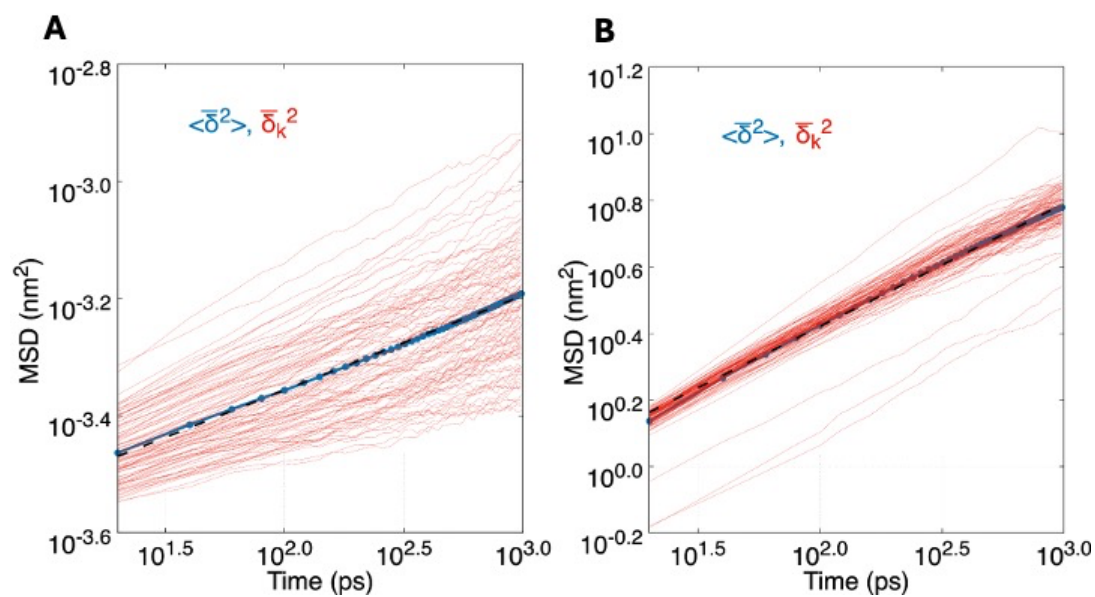

**Figure 2:** TA-MSDs $\overline{\delta_k^2(T,t)}$ (red lines) for each simualted replica and the average ensemble TA-MSD $\langle \overline{\delta^2(T,t)} \rangle$ (dotted blue line) for Villin(**A**) and CAP (**B**). The dashed black line is just a guide to the eye.

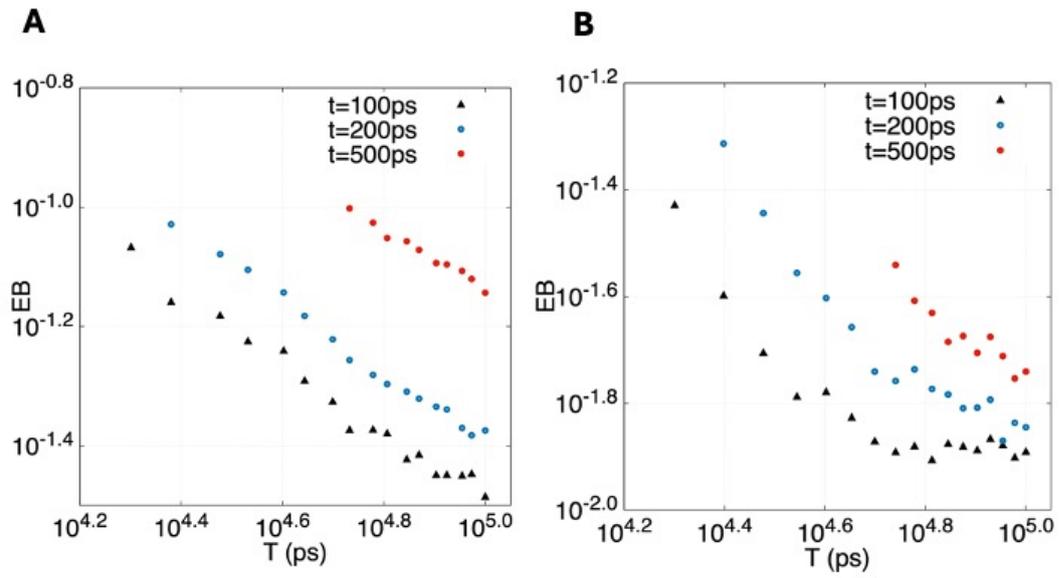

**Figure 3:** The $EB$ parameter varying $T$ and the lag time $t$ for Villin (**A**) and CAP (**B**).

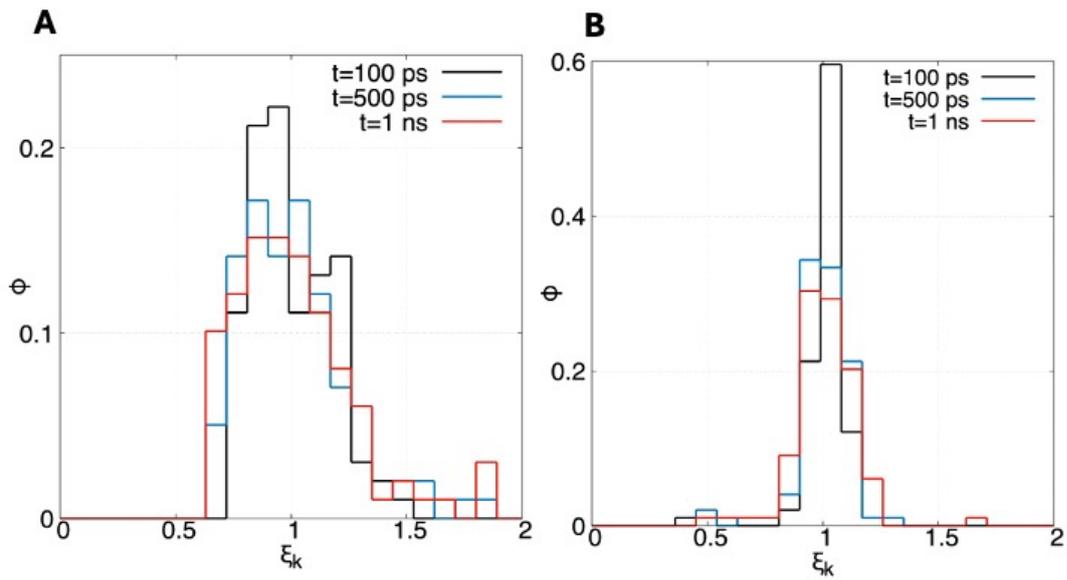

**Figure 4:** ϕ distribution varying the the lag time $t$ for Villin (**A**) and CAP (**B**)

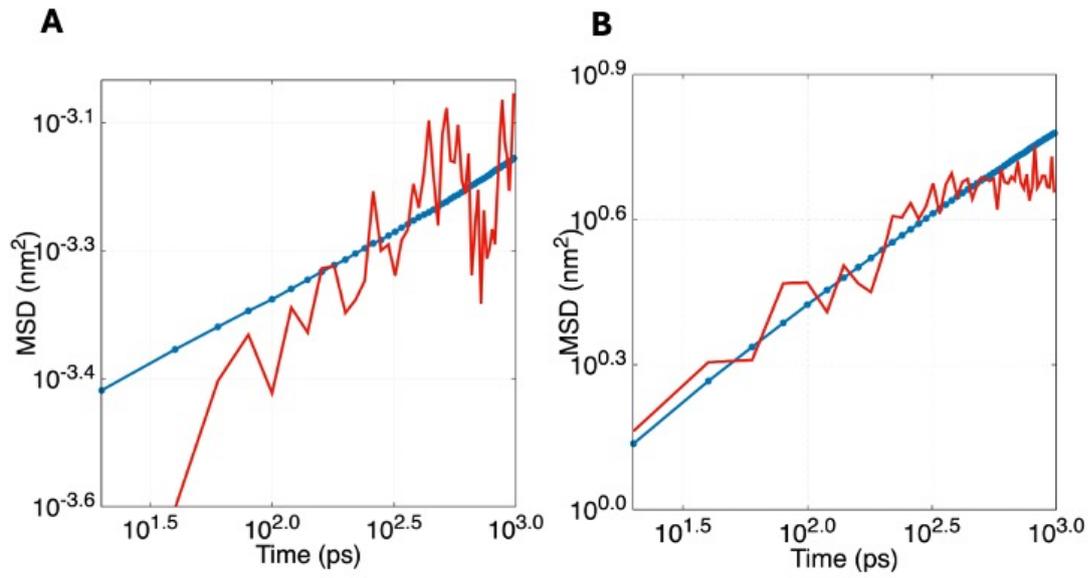

**Figure 5:** ensemble averaged MSD $\langle \Delta X^2(t) \rangle$ (red line) and ensembled avaregd TA-MSD $\langle \overline{\delta^2(T,t)} \rangle$ (botted blu line) for Villin (**A**) and CAP (**B**).